# Signature of Coherent Transport in Epitaxial Spinel-based Magnetic Tunnel Junctions Probed by Shot Noise Measurement

Takahiro Tanaka, Tomonori Arakawa, Kensaku Chida, Yoshitaka Nishihara, Daichi Chiba, Kensuke Kobayashi*, Teruo Ono, Hiroaki Sukegawa[1], Shinya Kasai[1], and Seiji Mitani[1]

*Institute for Chemical Research, Kyoto University, Uji, Kyoto 611-0011, Japan*
[1]*National Institute for Materials Science, Tsukuba, 305-0047, Japan*

We measured the shot noise in fully epitaxial Fe/MgAl$_2$O$_X$/Fe-based magnetic tunneling junctions (MTJs). While the Fano factor to characterize the shot noise is very close to unity in the antiparallel configuration, it is reduced to 0.98 in the parallel configuration. This observation shows the sub-Poissonian process of electron tunneling in the parallel configuration, indicating the coherent tunneling through the spinel-based tunneling barrier of the MTJs.

The tunnel magnetoresistance (TMR) effect in magnetic tunneling junctions (MTJs) has been one of the central topics in the spintronics field, not only because MTJs offer us an ideal stage to address spin-dependent transport, but also because MTJs have enormous potential in various applications[1]. In 2004, a very large TMR ratio was achieved in MTJs with a crystalline MgO barrier[1,2,3], where the coherent tunneling of highly spin-polarized electrons in the $\Delta_1$ state is considered to be essential[4]. More recently, a similarly large TMR ratio was obtained in the MTJs with a crystalline spinel (MgAl$_2$O$_4$) barrier[5]. The MTJs with a spinel barrier has two advantages, that is, its nondeliquescence and its small lattice mismatch (less than 1% for an Fe electrode) compared with that of the MgO barrier case (3-5 %). It is of primary importance to investigate whether the observed large TMR ratio is attributed to the coherent tunneling as well as in the MgO-based MTJs. Although conventional *I-V* characteristic measurements have been applied to address the transport properties in this system[5], additional experimental probes such as shot noise[6] would be preferable to clarify more the mechanism of electron tunneling.

In our previous paper[7], we reported the sub-Poissonian shot noise in MgO-based MTJs, indicating the coherent nature of the electron tunneling. Shot noise occurs when the current $I$ is injected into a tunnel junction. At zero temperature, the shot noise $S_I$ is expressed as $S_I = 2eIF$ with the Fano factor $F$. In the conventional tunneling process as in the normal metal/insulator/normal metal structure, $F=1$ holds[8], which means that each tunneling event through the barrier is independent ("Poissonian process"). In contrast, in MgO-based MTJs, $F$ was found to be less than unity in the parallel (P) confiuration, while $F = 1$ in the antiparallel (AP) configuration[7]. This observation is quantitatively consistent with the recent theoretical study by using the first-principles calculations[9], and proves the coherent electron tunneling through the MgO barrier.

Here, by using the same technique, we report the shot noise in MTJs with a crystalline spinel-based barrier. We observed a reduced Fano factor ($F=0.98 \pm 0.01$) in the P configuration, while we observed the full shot noise in the AP configuration. We also present 1/$f$ noise in the MTJs and compare it with MgO-based MTJs[10-16].

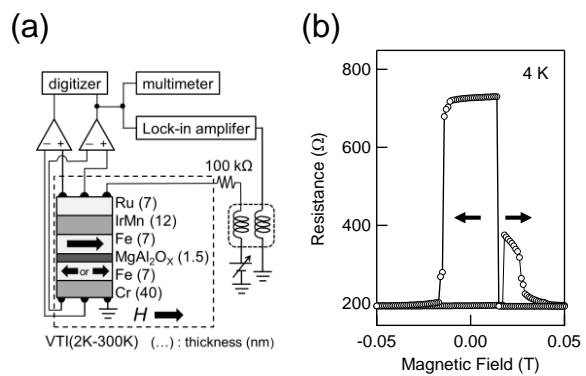

**Figure 1** (a) The present MTJs consist of Ta (3)/Au (120)/Cr (40)/Fe (30)/MgAl$_2$O$_X$(1.5)/Fe (7)/IrMn (12)/Ru (7)/Ta (3)/Au (120) (thickness in nm) multilayer. The measurement setup for the differential resistance and noise is schematically shown. The arrow shows the direction of the magnetic field. (b) MR curve of Sample B34 at 4 K.

The present MTJs consist of the multilayer stack of Cr (40)/Fe (30)/Mg (0.45)/Mg$_{33}$Al$_{67}$ (0.9)-O$_X$/Fe (7)/IrMn (12)/Ru (7) (thickness in nm) grown by magnetron sputtering on MgO(001) substrate [Fig. 1(a)]. The (Mg/Mg-Al)-O$_X$ barrier layer was fabricated by plasma oxidation in an Ar + O$_2$ atmosphere, and the final barrier thickness was around 1.5 nm. The multilayer from the Cr buffer to the top Fe layer grew epitaxially with a high (001) orientation. Details of the multilayer fabrication process and structural characterization will be reported elsewhere[17]. The multilayers were patterned into 1.5×0.5 μm$^2$ by photolithography and Ar ion milling after annealing at 175°C for 15 min under a magnetic field of 5 kOe. The measured two samples (Samples B32 and B34) give consistent results between each other as described below. Figure 1(b) shows a typical magnetoresistance (MR) curve for B34 at 4 K. For B34, the MTJ resistances in the P and AP configurations ($R_P$ and $R_{AP}$) are 192 and 730 Ω, respectively, and the MR ratio defined by ($R_{AP}$-$R_P$)/$R_P$ is 280 % with the area resistance (RA) of 144 Ω μm$^2$ [see Table I for more details]. The MR ratio of B32 is smaller than that of B34, which might be due to the incomplete AP configuration in B32.

*E-mail address: kensuke@scl.kyoto-u.ac.jp

**TABLE I.** List of Fano factors and Hooge parameters for both configurations with the MR ratios and RAs at 4 K.

| Sample No. | Fano factor | Hooge Parameter (μm²) | MR ratio (%) | RA (Ωμm²) |
|---|---|---|---|---|
| B34 (P) | 0.979±0.009 | $1\times10^{-12}$ | 280 | 144 |
| B34 (AP) | 1.002±0.010 | $2.5\times10^{-11}$ | | |
| B32 (P) | 0.981±0.008 | $1.3\times10^{-12}$ | 190 | 143 |
| B32 (AP) | 1.01-1.05* | $2.5\times10^{-10}$ | | |

The noise measurements[7]) were carried out in the variable temperature insert (Oxford) as schematically shown Fig. 1(a). The dc current with a small ac modulation is applied to the MTJ through a 100 kΩ resistor to obtain the differential resistance via the standard lock-in technique. Two voltage signals across the MTJs are amplified independently using two amplifiers (NF Corporation LI-75A) placed at room temperature and recorded with a two-channel digitizer (National Instruments PCI-5922). In order to reduce the external noise, the measured two sets of time domain data are cross-correlated to yield the noise power spectral density through the fast Fourier transformation. We performed the histogram analysis[16]) of the measured voltage noise power spectral density $S_V$, typically between 140 and 180 kHz (4000 points). By carefully calibrating the measurement system and statistically treating the errors, we are able to determine the Fao factor well within the accuracy of 1%.

Figures 2(a) and 2(c) show the experimental result of the differential resistance (d$V$/d$I$) and the current noise power spectral density $S_I$ for Sample B34 at 4 K as a function of the bias voltage ($V_{sd}$) for the P and AP configurations, respectively. The differential resistances in both configurations show the asymmetric zero-bias peak structure, which is consistent with the previous report for this type of MTJs[5]). $S_I$ is obtained from $S_V = (dV/dI)^2 S_I$. At $V_{sd} = 0$, $S_I$ is equal to the thermal noise $4k_BT$ d$V$/d$I$ with $T = 4$ K. The obtained $S_I$ is symmetric to the bias-voltage reversal. The parabolic behavior at finite bias ($|eV_{sd}| \sim k_BT$) indicates the crossover from the thermal to shot noise, and $S_I$ is linearly dependent on $V_{sd}$ for $|eV_{sd}| \gg k_BT$. All these features agree with those expected from the conventional shot noise theory.

In order to obtain $F$, the numerical fitting is performed by using the following equation,

$$S_I = 4k_BT/\frac{dV}{dI} + 2F\left[eI\coth\left(\frac{eV_{sd}}{2k_BT}\right) - 2k_BT/\frac{dV}{dI}\right].$$

The numerical fitting was performed by using the nonlinear least-square regression analysis taking the experimental errors into account. The results of the numerical fitting are superposed in Figs. 2(a) and 2(c) for the P and AP configurations, respectively. The solid and dashed curves represent the fitted curve and the Poissonian case ($F$=1),

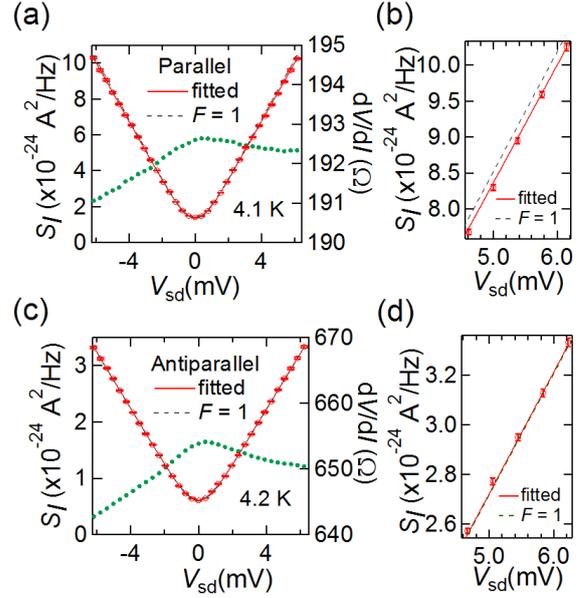

**Figure 2** (a) Differential resistance (solid circle) and current noise power spectral density (open circle) of B34 for the parallel configuration obtained from the histogram analysis between 140 and 180 kHz. The solid line is the fitting curve and the dashed line represents the curve corresponding to $F = 1$. (b) Part of Fig. 2(a) is zoomed to show that experimental data is surely deviated from the dashed line. (c) and (d) Counterpart of Figs. 2(a) and 2(c) for the AP configuration, respectively.

respectively. Figures 2(b) and 2(d) show the corresponding expanded view of Figs. 2(a) and 2(c). As seen in Fig. 2(b), $F$ for the P configuration ($F_P$) is reduced from unity to be $F$=0.979±0.009, where the error bars indicate 99.73% confidence interval. The Fano factor and MR ratio of Samples B32 and B34 are compiled in Table I. In both cases, the $F_P$ is less than 1, while $F$ for the AP configuration ($F_{AP}$) is statistically equal to 1.00 for B34 [see Fig. 2(d)]. We note that in Sample B32 in the AP configuration, the 1/$f$ noise contribution is not negligible, and the estimated factor ranges between 1.01 and 1.05 depending on the frequency range for the analysis, although the 1/$f$ noise is confirmed to affect no influence on the $F_P$ value for B32. The observation of the finite reduction of $F_P$ from unity is the central result of the present experimental work.

Previously, we reported the Poissonian shot noise in MgO-based MTJs with the 1.5-nm -thick barrier[16]. Later, we also reported spin-dependent suppression of the Fano factor (typically 0.91 in the P configuration and 0.99 in the AP configuration) in MgO-based MTJs with a barrier as thin as 1.05 nm[7]). These results are in agreement with the recent theoretical study[9]), where the shot noise in disordered Fe/MgO/Fe tunnel junctions was calculated from first principles. Thus, the observed sub-Poissonian shot noise in the MgO-based MTJs gives the evidence for coherent electron tunneling in the MgO barrier. Although the coherent transport via the $\Delta_1$ states has been inferred theoretically and experimentally, a convincing experimental signature can be obtained through the shot noise measurement.

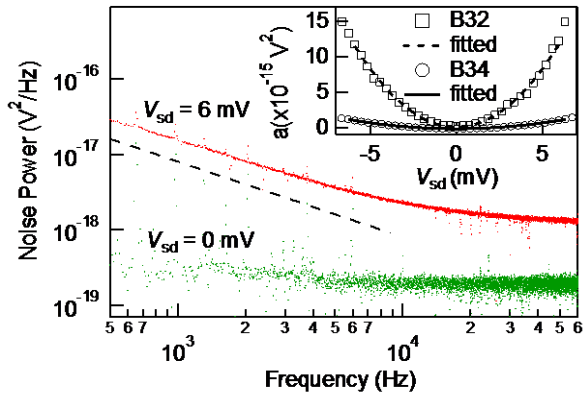

**Figure 3** Magnitude of 1/*f* noise for the AP configuration with applied voltage at 4 K. the square shows B32 and the dashed line is its fitting. The circle and solid line are counterparts of B34.

The present result of the shot noise strongly indicates that the above story is also applicable here, as the relevance of the coherent tunneling of $\Delta_1$ states in the spinel-based MTJs was theoretically discussed very recently[18]. The observed sub-Poissonian process in the P configuration indicates that the electron tunneling event through the spinel barrier is anticorrelated with each other due to the coherent tunneling. On the other hand, it should also be noted that the barrier thickness of the present MTJs is as thick as 1.5 nm, where the Poissonian shot noise would be the case theoretically for MgO-based MTJs[9]. At this moment, we do not understand the exact reason why the suppression of $F_P$ is observed for a thick case, while it may indicate the difference in the tunneling process between spinel and MgO-based MTJs. Further theoretical treatment and systematic experimental work are necessary to clarify this problem.

As a remark, the experimental observation that $F_{AP} > F_P$ with $F_{AP}$ significantly suppressed below 1 was reported for amorphous $Al_2O_3$-based MTJs[14], which was attributed to the tunneling mediated by localized impurity states inside the barrier. However, the present result of $F_{AP} > F_P$ with $F_{AP}$ very close to 1 is not likely to be explained within this model, unless unrealistic parameters regarding the barrier properties and the spin polarization are assumed (see Fig. 5 in Ref. [14]).

Finally, we discuss the 1/*f* noise in our device. From the obtained spectral density between 6 and 60 kHz, we derive the 1/*f* noise as a function of $V_{sd}$. The result for the AP configuration is shown in the inset of Fig. 3. The 1/*f* noise is expressed as $a/f$. As is usually the case for the 1/*f* noise, the factor *a* shows a parabolic behavior as a function of $V_{sd}$. The Hooge parameter is defined as $\alpha = aA/V_{sd}^2$, where *A* is the junction area ($\mu m^2$). The values of α are summarized in Table I. For the P configuration, α is much smaller than the AP configuration, which was also the case in the MgO-based MTJs[16]. It may indicate that the origin of 1/*f* noise for the P configuration is charge traps, whereas the considerable 1/*f* noise contribution in the AP configuration suggests the magnetic origin[16]. The Hooge parameter of the high-quality MgO-based MTJs[7] was 3.4 x $10^{-13}$ $\mu m^2$ at 3 K for the P configuration. Although the sample geometries are not exactly the same, this value is roughly comparable to those of the present spinel-based MTJs. Such small 1/*f* noise contribution indicates the high quality of the well-crystalized spinel barrier.

In conclusion, the shot noise is measured in Fe/$MgAl_2O_4$/Fe-based MTJs. The reduced Fano factor (0.98) is observed for the P configuration, indicating the sub-Poissonian process of the electron tunneling, while the Poissonian shot noise is obtained for the AP configuration. This observation strongly suggests the relevance of the coherent transport in this type of MTJs.

This work was partially supported by the JSPS Funding Program for Next Generation World-Leading Researchers.